\documentclass[twocolumn,showpacs,preprintnumbers,amsmath,amssymb]{revtex4}

\usepackage{graphicx}
\usepackage{dcolumn}
\usepackage{bm}

\def\beq{\begin{equation}}
\def\eeq{\end{equation}}

\begin{document}

\title{Proton Decay and Sub-Structure}

%

\author{Kazuo  Koike}
 \email{koike@ed.kagawa-u.ac.jp}
\affiliation{%
Department of Natural Science, Kagawa University,
 Takamatsu 7608522, Japan\\
}

\date{\today}

\begin{abstract}
The problem of elementarity of electric charge is investigated.
It is known that possible proton decay is caused by $X$ and $Y$ bosons
which are characteristic in GUTs. The charge structure of GUTs is, 
however, not simple, and the five types of charge units appear. We investigate
the problem of elementarity in taking into account of the simplest model of the 
charge structure known as the rishon model, where the charge units are known as 
1/3 ($T$) and 0 ($V$). 
In order to approach to sub-structure together with
the problem of simplicity of charge structure, possible rearrangements of rishons 
are investigated on the basis of color and hypercolor confinement condition. 
Further, possible existence of the simplest electroweak sub-structure is discussed.
Our model predicts possible proton decay mode, in which  $X$ and $Y$ bosons are not
concerned.

\end{abstract}

\maketitle

\section{Introduction}

In the progress of high-energy physics, so-called energy frontier have reached
to a few TeV region, where discoveries of new phenomena
are expected.
In such situation, it is well known that the five unites of electric charge,
 \beq
Q ~=~~0,~~~~1/3,~~~~2/3,~~~~1,~~~~4/3 
\eeq
\noindent
appear in GUTs\cite{rf:SU5},
where the 4/3 charge is carried by $X$ gauge boson, which is considered to be the 
particle responsible for proton decay.

Thus, so far as the charge units are concerned, it seems that GUTs are too 
complicated to be the final theory of elementary particles.
Especially, in any gauge model such as GUTs, the $\rm U(1)$ symmetry concerning 
to the conservation of electric charge is absolutely maintained in the final stage
of steps of symmetry breaking. 
Furthermore, it is known that the electric charge remains and is conserved even in a 
critical condition such as the gravitational collapse of matter, where almost all 
quantum numbers looses their meanings\cite{rf:blackhole}.
Thus, the electric charge seems to be one of the most fundamental quantities in 
particle physics.
It is known that the simplest model based on the elementarity of electric charge
is the rishon model\cite{rf:rishon} of leptons and quarks.
In this model,the leptons and quarks are given as
\beq
        { u =  TTV,~~
          d =  \tilde{V}\tilde{V}\tilde{T},~~}
          {\nu} = {VVV,~~
          e^- = \tilde{T}\tilde{T}\tilde{T}}
          \label{eq:Rishon}
\eeq
where the $T$ rishon has the electric charge $e/3$, while the $V$ rishon is purely 
electrically neutral.
 
Though models of leptons and quarks are, however, in the stage of schematical model,
some remarkable minimal dynamical approaches are proposed\cite{rf:Harari-Seiberg}.
For further steps, it is possible to treat schematically possible processes of 
sub-constituents under minimal dynamical assumptions for composite system.
On the basis of such approach, we will investigate possible existence of new processes,
and possible dynamical approach to sub-system of leptons and quarks.
This paper is concerning to theses problems. 

We will assume that leptons and quarks are composite system of sub-constituents 
confined with the hypercolor singlet conditions, 
as have been assumed in dynamical rishon mode. 
In our schematical approach, it is pointed out that possible existence of new proton 
decay mode, which is different from that of GUTs, is suggested.


\section{Color and hypercolor confinement condition}

We will follow the policy of so-called dynamical rishon model\cite{rf:Harari-Seiberg},
where it is assumed that the hypercolor singlet states of composite system are 
realized as ordinary particles such as leptons and quarks.

Though the rishon is assumed to be Dirac particle in the dynamical rishon model,
while we have proposed, under the development of Majorana particle model of neutrinos,
Majorana V rishon model\cite{rf:Koike}. In this paper, however, we will at first  
restrict ourselves to Dirac particle model, and the case of Majorana model will be 
investigated in later.

 In the dynamical rishon model\cite{rf:Harari-Seiberg}, the exact symmetry realized by 
the local gauge group,
\beq
SU_3(C) \times SU_3(H)
\label{eq:CxH}
\eeq
is assumed for rishons, where C and H represents color and hypercolor, respectively.
The T rishon has  $Q=1/3$ and transforms like a $(3,3)$ representation of
$SU_3(C) \times SU_3(H)$.
The V rishon has  $Q=0$ and is assigned to $(\bar{3},3)$.
The $\tilde{V}$ or $\tilde{T}$ is in $(3,\bar{3})$ and $(\bar{3},\bar{3})$, 
respectively.

\section{Rearrangement Picture of sub-constituents under Confinement}

\begin{figure}
\includegraphics[scale=1.125]{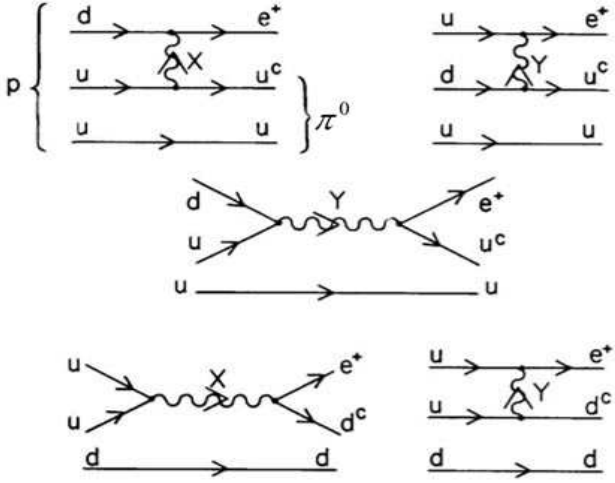}
\caption{\label{fig:P-GUT_all-1} Possible proton decay modes in GUTs. The proton decay
is caused by X and Y bosons. In sub-constituent level, the X or Y boson corresponds
to the configuration $\rm X  \sim (TTTTVV)$ , $\rm Y  \sim (TT\bar{T}V\bar{V}\bar{V})$,
respectively.
}
\end{figure}

We will treat for simplicity the simplest case of one generation,
and at first treat the case of Dirac type V rishon case. 
Some characteristics and problems
appearing in possible Majorana type V rishon model will be discussed in later. 

In the dynamical rishon model, the sub-constituents T and V are supposed to 
be confined in leptons and quarks by hypercolor confinement mechanism 
realized by the hypercolor singlet condition.
Under such condition, the composite system exhibits various rearrangement processes.
The electroweak structure supposed to appear as a result of such 
mechanism\cite{rf:effective}.
In this paper, we will investigate rearrangements of sub-constituents
under the confinement condition, together with possible sub-electroweak structure. 
It should be emphasized that in the rearrangement of sub-constituent, 
possible existence of new mode of 
proton-decay is suggested, which is different from that of GUTs case where the 
proton-decay is mediated by the $X^{\alpha}$ and $Y^{\alpha}$ gauge bosons with 
the color index $\alpha$.

The TTTVVV configuration belong to $SU_3(C) \times SU_3(H)$ singlet, and is
identified to $W^+$ boson. In the same way, the GUTs gauge bosons
$X^{\alpha}$ and $Y^{\alpha}$ with color index $\alpha$ corresponds to the configuration 
TTTTVV and $\rm TTV\bar{T}\bar{V}\bar{V}$, respectively, and they belong to $(3,1)$ 
representation of of $SU_3(C) \times SU_3(H)$.
In our sub-constituents model, it should be emphasized that the process caused by 
exchange of the composite state with ${\rm T\bar{V}}$ configuration is also expected. 
The ${\rm T\bar{V}}$ belongs 
to $(3,1)$ representation, similar to the case of $X^{\alpha}$ and $Y^{\alpha}$ bosons.

It is well-known that the exchange of $X^{\alpha}$ and $Y^{\alpha}$ bosons
causes possible proton-decay in GUTs, as is shown in FIG.\ref{fig:P-GUT_all-1}.
Then, what processes are caused by 
the exchange of composite state of ${\rm T\bar{V}}$?

\section{ A new proton-decay mechanism originates in sub-structure} 


 It is known that the X boson with electric charge 4/3 and Y boson with charge 1/3 
mediate the proton decay process
in GUTs such as $\rm SU(5)$ model. The boson X or Y correspond to TTTTVV , 
$\rm TTV\bar{T}\bar{V}\bar{V}$ configuration, respectively.


\begin{figure}
\includegraphics[height=2.5cm,width=7.5cm]{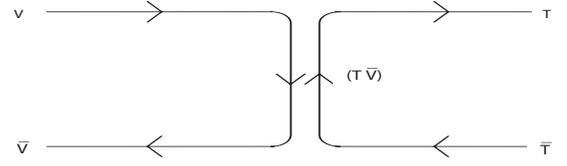}
\caption{\label{fig:TV_element-2}An elementary rearrangement in sub-system.
}
\end{figure}

It should be emphasized, however, the new type of proton decay mode appears, 
which is mediated by the exchange of ${\rm T\bar{V}}$ pair with charge 1/3, 
in a composite model of leptons and quarks.
The elementary process mediated by the rearrangement of sub-constituents, 
given in FIG.~{\ref{fig:TV_element-2}},

\beq
V + \tilde{V} \to T + \tilde{T}
\label{eq:elementary}
\eeq

\noindent
will cause the processes
\beq
(TTV) \to (TTT)~~,~~
(\tilde{V} \tilde{V} \tilde{T}) \to (\tilde{V} \tilde{T} \tilde{T})
\label{eq:induced}
\eeq

\noindent
which are represented as

\beq
u \to e^+ ~,~~
d \to u^c
\label{eq:induced2}
\eeq

\noindent

\noindent
\noindent

\begin{figure}
\includegraphics[scale=0.6]{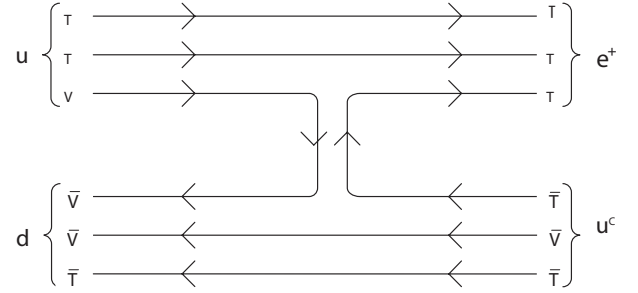}
\caption{\label{fig:TV_f0-3} Possible $ u + d \to e^+ + u^c$ process through the 
exchange of $(\rm T\bar{V})$ pair.
}
\end{figure}

\noindent
The explicit rearrangement diagram of the process in Eq.~(\ref{eq:induced2})
is given in FIG.~\ref{fig:TV_f0-3}.
Thus, the transition of composite quarks, which is caused by this rearrangement

\beq
(uud) \to (e^+ u^c u)~
\label{eq:ProtonDecay0}
\eeq

\noindent
is induced. This is just the proton decay process,

\beq
p \to e^+  +~ {\pi}_0
\label{eq:ProtonDecay}
\eeq

\noindent
mediated by the exchange of ${\rm (T\bar{V}})$ pair.
That is, in our model the proton decay is caused without the X or Y gauge bosons, 
which are characteristic in GUTs as shown in FIG.~\ref{fig:P-GUT_all-1}.

\section{The effective and possible fundamental interaction} 

\begin{figure}
\includegraphics[height=2.35cm,width=7.5cm]{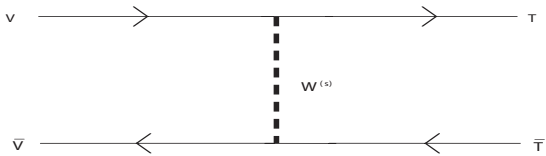}
\caption{\label{fig:TV_element-4}Elementary process in sub-system caused by the 
exchange of effective sub-gauge boson $W^{(s)}$ .
}
\end{figure}

The elementary process in Eq.~(\ref{eq:elementary}) will be represented
as FIG.~\ref{fig:TV_element-2}. This process will also represented as an effective 
process mediated by the effective boson $W^{(S)\pm}_\alpha$ with color indexes 
$\alpha$, as shown in FIG.~\ref{fig:TV_element-4}.

\noindent
We will introduce an effective interaction of this process,
and at the same time, we will investigate possible scenario to identify it as the
fundamental one.
For simplicity, we will restrict to the one generation case in this paper. 
Assuming the weak-doublet in sub-level as,

\beq
        \Psi_L =  { T \choose  V^c}_L
          \label{eq:Rishon2}
\eeq

\noindent
the effective electroweak Lagrangian in our system is given as

\begin{eqnarray}
\nonumber\\
     {\cal L} =  &-& ~\bar{\Psi}_L \gamma_\mu(\partial_\mu - ig\mathbf{t}\cdot\mathbf{A^{(s)}_{\mu}} -ig^\prime YB^{(s)}_{\mu} )\Psi_L 
\nonumber\\
 &-& \sum_{\kappa=T,V}~\bar{\Psi}_R^{\kappa }(\partial_\mu -ig^\prime YB^{(s)}_\mu){\Psi}_R^{\kappa }
\nonumber\\
&-& ~(\bar{\Psi}_L \phi {\Psi}_R^{V } M^{(V)} + h.c.)
\nonumber\\
&-& (\bar{\Psi}_L(-\eta)\phi^* {\Psi}_R^{T} M^{(T)} + h.c.)
\label{eq:EW_Sub}
\end{eqnarray}

\noindent
where $\mathbf{A}^{(s)}_\mu $ and $B^{(s)}_{\mu}$ are the gauge fields concerning 
to sub-system.
The electric charge of component of doublet in Eq.~(\ref{eq:Rishon2}) is given as
\beq
Q = I_3  +  Y
\label{eq:charge_sub}
\eeq

\noindent
in the unit of e/3, where $Y = -1/2$.
After symmetry breaking, the sub-gauge bosons $W^{(S)\pm }_\mu$,  $Z^{(S)}_\mu$ and $A^{(s)}_\mu$ bosons  appear, where $W^{(S)\pm }_\mu$ is characterized by the 
charge unit e/3.
The ordinary electroweak gauge bosons $W_{\mu}$, $Z_{\mu}$ and $A_{\mu}$ will be 
symbolically expressed by making use of the sub-gauge bosons $W^{(s)}_{\alpha\mu}$, $Z^{(s)}_{\mu}$ and $A^{(s)}_{\mu}$ as,

\begin{eqnarray}
W_{\mu} &=& ({\epsilon^{\alpha\beta\gamma}W^{(s)}_\alpha W^{(s)}_\beta W^{(s)}_\gamma)}_{\mu}~\nonumber\\
Z_{\mu} &=& Z^{(s)}_{\mu}~\nonumber\\
A_{\mu} &=& A^{(s)}_{\mu}
\label{eq:effective_EW}
\end{eqnarray}
\noindent
where $W_{\mu}$ is $SU_3(C) \times SU_3(H)$ singlet, while
$W^{(s)}_\alpha$ belong to $({3},1)$ representation. 

It should be noted that $A^{(s)}$ and $Z^{(s)}$ are realized as singlet as it is, 
provided that the correspondences
$A^{(s)} \sim (\bar{T}T)$ and $Z^{(s)} \sim (\bar{V}V)$ together with an appropriate 
mixing are assumed.
The Eq.~(\ref{eq:effective_EW}) is a symbolical representation of
composite system based on yet to be known mechanism.
The sub-gauge bosons $W^{(S)}_\mu$ should be confined in the ordinary gauge bosons 
belonging to the level of leptons and quarks.
The effective electroweak structure in the level of leptons and quarks will be induced 
on the basis of Eq.~(\ref{eq:effective_EW}).

In order to investigate the properties of composite system given by 
Eq.~(\ref{eq:effective_EW}), let us examine the familiar (inverse) $\beta$-decay process 
of d quark,

\beq
d + {\nu}_e    \to  u + e^- 
\label{eq:beta}
\eeq

\noindent
In the picture of fundamental constituents, Eq.~(\ref{eq:beta}) is rewritten as,

\beq
(\tilde{V}\tilde{V}\tilde{T}) + (VVV) \to (TTV) + (\tilde{T}\tilde{T}\tilde{T})
\label{eq:beta-rishon}
\eeq

\noindent
Thus, we see that the fundamental interaction in our process is given as

\beq
{\cal L} = g_{w}^{(s)} \bar{T}\gamma^\mu (1 + \gamma_5 ) V^c W_\mu^{(s)}
\label{eq:fundamental_int}
\eeq

\noindent
and corresponding current is
\beq
{\cal J} = \bar{T}\gamma^\mu (1 + \gamma_5 ) V^c
\label{eq:fundamental_current}
\eeq

\noindent
which belongs to $({3},1)$ representation. The interaction in the process
given by Eq.(\ref{eq:elementary}), 
\beq
\nonumber
V + \tilde{V}  \to T + \tilde{T}
\eeq

\noindent
which is caused by only $({3},1)$ single current, should be extremely suppressed.

On the contrary, in our scheme, the ''composite" process with 3-fold single elementary 
processes such as Eq.(\ref{eq:beta-rishon}) 
should be not so suppressed. The single elementary process in 
Eq.(\ref{eq:elementary}) is expected to be extremely suppressed 
in the same way as the 3-body quarks are almost completely confined in hadrons and 
realization of colored configuration is extremely suppressed. 
The process such as Eq.(\ref{eq:beta}) or equivalently  
Eq.(\ref{eq:beta-rishon}) are enhanced, similar to the realization of the 3-body 
color-singlet states of quarks are enhanced. That is, the process given in 
Eq.(\ref{eq:beta-rishon}) is essentially the non-perturbative process with extreme 
suppression of non-color-singlet currents. Thus, we will investigate possible scheme 
based on fundamental interaction.

\section{Possible approach on the basis of the fundamental interaction in sub-system}

We have introduced the effective interaction corresponding 
to the rearrangement picture in sub-system. We will investigate here the possible
case that Eq.(\ref{eq:EW_Sub}) is not an effective interaction but the 
most fundamental one.
That is, it is worth while to investigate the possibility that the fundamental 
interaction cause various effective interactions such as the standard model itself, 
under the restriction of exact symmetry $SU_3(C) \times SU_3(H)$ given in 
Eq.(\ref{eq:CxH}). 
In appearance of effective interaction, especially, color and hypercolor
confinement condition will carry most important role. It should be noted that this 
approach is partially similar to the case of meson theory for nuclear force where 
various properties of nucleus are derived on the basis of QCD.

It should be emphasized that under development along such strategy it may become 
possible to resolve the hierarchy problem on coupling constants and masses, which
are only given as parameters in the standard model or GUTs.

\section{ Remarks on the case of Majorana V rishon} 

In progress of neutrino physics, the existence of Majorana particle seems to be almost 
confirmed.
In addition to the neutrinos, are there any Majorana particle in nature?
In such situation, we have proposed in a few years ago a model of leptons and 
quarks on the basis of possible Majorana V theory together with possible existence of 
its partner in the framework of rishon model\cite{rf:Koike}. 

We will investigate possible representation of sub-constituent in our Majorana V model,
under the local gauge group $SU_3(C) \times SU_3(H)$.
In the simplest $SU_3(H)$ hypercolor model, the assignment to realize
the hypercolor singlet states is rather complicated\cite{rf:Octanion}.
The T rishon is assigned to $(3,8)$ representation of
$SU_3(C) \times SU_3(H)$.
The V rishon is assigned to $(8,8)$.
The $\tilde{V}$ and $\tilde{T}$ are, respectively, in $(8,8)$ and $(\bar{3},8)$.

The T rishon has  $Q=1/3$ and transforms like a $(3,8)$ representation of
$SU_3(C) \times SU_3(H)$.
The V rishon has  $Q=0$ and is assigned to $(8,8)$.
The $\tilde{T}$ corresponds to $(\bar{3},8)$ representation.
For Majorana type $V$, the $\tilde{V}$ is equivalent to $V$ and also assigned to $(8,8)$.
The quarks u and d belong to $(\bar{3},1)$ representation.

It should be noted that the results given in the previous sections will be almost
maintained in Majorana V model, except the appearance of characteristic picture of
Majorana particles that the fermion number is not conserved.
Furthermore, the restrictions due to color and hypercolor confinement will be rather 
weakened, then possible existence of some new particles are expected.


\section{ Discussion and Summary} 

In this paper, we have investigated possible existence of new proton decay mechanism
on the basis of a rearrangement picture of sub-system of leptons and quarks, in
which the sub-constituents are confined in leptons and quarks by the color and
hypercolor confinement condition based on $SU_3(C) \times SU_3(H)$ group.
 
We have discussed the possibility that the pre-electroweak structure 
appears in the rishon level, and a certain kind of confinement of 
pre-gauge bosons realizes the well-known electroweak structure in the level 
of leptons and quarks.  In our approach, it may be possible to
calculate the magnitude of coupling constants and the masses of composite
leptons and quarks, which are given as parameters in pure field theoretical approaches.
We have assumed in this paper that the gauge structure of the Standard Model appears 
as a result of the dynamics of the sub-levels of leptons and quarks.

Is the sub-system of leptons and quarks completely subject to quantum mechanics?
It should be noted that the standard formulation of the rishon model is not yet known, 
and it is probable that the rishon behaves as if it is partially the 
objects beyond the ordinary quantum field theory\cite{rf:Maki},\cite{rf:Blokhin}. 
In order to approach to such system, however, it 
will be meaningful to treat it in the framework of conventional field theory 
with color and hypercolor confinement.


Finally, is there really a sub-level below the level of leptons and quarks? 
In our opinion, such a sub-level surely exists, and disclosing it will make it 
possible to predict theoretically quantities such as the Higgs coupling constants,
the magnitudes of symmetry breaking
$\langle\phi\rangle$, the mass spectrum and mixing parameters of all particles, etc.
We conclude this paper with the note that we have restricted ourselves in this 
paper to the proposal of the framework of possible approach to sub-system,
and details  will be discussed in near future.

\section*{Acknowledgments}
 This work was supported by Grant-in-Aid of Japanese Ministry of 
 Education, Science, and Culture (15540384).



\begin{thebibliography}{9}
\bibitem{rf:effective}
Possible realization of electroweak structure is also discussed in the 
dynamical rishon model. See ref\cite{rf:Harari-Seiberg}.
\bibitem{rf:Octanion}
It should be noted that formulations based on no associated algebra
are also possible.
\end{thebibliography}
\end{document}